\documentclass[conference]{IEEEtran}
\IEEEoverridecommandlockouts
\usepackage{cite}
\usepackage{amsmath,amssymb,amsfonts}
\usepackage{algorithmic}
\usepackage{graphicx}
\usepackage{textcomp}
\usepackage{xcolor}
\usepackage{subcaption}
\usepackage{url}
\def\BibTeX{{\rm B\kern-.05em{\sc i\kern-.025em b}\kern-.08em
    T\kern-.1667em\lower.7ex\hbox{E}\kern-.125emX}}
\begin{document}
\title{Complex VAE with Heavy-Tailed Likelihood for Radar Target Detection in Sea Clutter}
\author{
	\IEEEauthorblockN{Ting Bai, Jun Tang, Yuxin Xu}
	\IEEEauthorblockA{\textit{dept. Electronic Engineering,}
		\textit{Tsinghua University,}
		Beijing, China}
}
\maketitle
\begin{abstract}
	To address the heavy-tailed, spike-prone nature of sea clutter and the scarcity of labeled target data, an unsupervised complex-valued variational autoencoder (VAE) for maritime radar target detection is proposed. In implementation, each complex baseband slow-time sequence is represented by its in-phase and quadrature components, and the model learns their joint reconstruction from clutter-only data. A Student-\(t\) negative log-likelihood is adopted to capture heavy-tailed reconstruction errors while reducing sensitivity to outliers during clutter learning. In addition, a time-domain amplitude error constraint is introduced to penalize slow-time magnitude mismatch in the reconstruction. At inference, reconstruction deviation is used as the detection statistic, and the decision threshold is set via an empirical quantile estimated from a clutter-only validation set to enforce a constant false-alarm rate (CFAR). Experiments on measured sea-clutter data show that detection performance is consistently improved over MF, AMF, and a real-valued \(\beta\)-VAE under CFAR constraints.
\end{abstract}

\begin{IEEEkeywords}
	Unsupervised learning, radar target detection, sea clutter, VAE.
\end{IEEEkeywords}

\section{Introduction}
Maritime surveillance radar is widely used for vessel traffic management, search and rescue, and situational awareness. However, weak targets can be severely obscured by strong sea clutter under heterogeneous sea conditions. Meanwhile, these targets often exhibit a low signal-to-clutter-plus-noise ratio (SCNR), which further complicates robust detection. Consequently, reliable weak-target detection in sea-clutter background remains a fundamental challenge in radar signal processing.

Classical radar detectors usually construct an explicit test statistic from secondary clutter samples and a prescribed background model. Examples include covariance-based matched filtering, such as the matched filter (MF)\cite{2007An} and adaptive matched filter (AMF)\cite{fuhrmann1992cfar}, together with regularized covariance estimation for radar clutter\cite{jain2023radar}. To better describe sea-clutter tails, K-distribution\cite{dong2012optimal}, log-normal\cite{xue2021adaptive}, and compound-Gaussian models\cite{guo2024persymmetric} have also been used in adaptive detection and threshold design\cite{wu2025adaptive}\cite{jiang2023subspace}. These methods are interpretable and well grounded, but their performance can degrade when the assumed clutter model or the available secondary data do not match the operating scene.
%When sea conditions change rapidly, frequent recalibration is typically required, thereby increasing engineering complexity.

Learning-based detectors offer another route by estimating target--clutter discrimination rules from data. Neural networks\cite{feintuch2023neural}, convolutional networks\cite{qu2022false}\cite{chen2021false}, sequence models\cite{wan2022sequence}, and attention mechanisms\cite{wang2022maritime} have been used to extract radar features and produce detection scores. Their practical use in maritime surveillance is constrained by the scarcity of reliable target labels, especially when the sea state, viewing geometry, and operating mode vary across datasets.

Unsupervised reconstruction methods are attractive in this setting because clutter-only measurements are much easier to collect than labeled target returns. In such methods, a generative model is trained on nominal clutter, and the reconstruction mismatch is later used as an anomaly score. Recent VAE-based radar detectors have demonstrated this idea in cluttered environments\cite{liang2022unsupervised}\cite{rouzoumka2025out}\cite{ru2025optimizing}. For sea-clutter detection, three implementation issues are particularly relevant. The I/Q components of the complex baseband slow-time signal should be reconstructed consistently, spike-like clutter samples should not dominate the training loss\cite{akrami2019robust}, and the reconstruction criterion should provide sufficient contrast for weak-target thresholding.

Motivated by the above observations, a reconstruction-based complex-valued variational autoencoder (VAE) is developed to improve the reliability of reconstruction-deviation-based detection statistics. In implementation, the complex baseband sequence is represented by its I/Q components, and the network learns their joint reconstruction from clutter-only data. The main contribution of this conference paper is a compact reconstruction-based detector that combines a Student-\(t\) likelihood with a time-domain amplitude error constraint. The former reduces the influence of spike-like clutter samples during background learning, while the latter encourages the reconstructed slow-time magnitude sequence to remain consistent with the target-free input.

Experiments are conducted on measured X-band sea-clutter scanning data, with MF, AMF, and a real-valued \(\beta\)-VAE included as baselines under the same CFAR thresholding protocol. The comparison and ablation results are used to evaluate the effect of the heavy-tailed likelihood and the amplitude reconstruction constraint.

\section{Signal Model}
For each cell under test (CUT), an \(L\)-pulse slow-time vector is formed from the complex baseband echoes. Under \(H_0\), the CUT contains sea clutter and receiver noise only; under \(H_1\), a moving point target contributes a Doppler-dependent component:
$$
	\begin{cases}
		H_0:\ \mathbf{x} = \mathbf{c} + \mathbf{n}, \\
		H_1:\ \mathbf{x} = \alpha \mathbf{p}(f_d) + \mathbf{c} + \mathbf{n},
	\end{cases}
$$
where \(\mathbf{x},\mathbf{c},\mathbf{n}\in\mathbb{C}^{L}\) denote the received vector, sea-clutter return, and thermal noise, respectively. The coefficient \(\alpha\in\mathbb{C}\) contains the target amplitude and initial phase. Over a short dwell, a target with constant radial velocity is represented by the steering vector
$$
	\mathbf{p}(f_d)=
	\left[
		1,\ e^{j2\pi f_d T},\ e^{j2\pi f_d 2T},\ \ldots,\ e^{j2\pi f_d (L-1)T}
		\right]^T,
$$
where \(f_d\) is the target Doppler frequency and \(T\) is the pulse repetition interval. The model is used to define the target-present test samples, while the VAE is trained using clutter-only samples associated with \(H_0\).

\section{Methods}
An unsupervised reconstruction-based framework is developed with separate training, validation, and testing datasets. During training, clutter-only \(H_0\) snapshots are represented as I/Q slow-time sequences and used to optimize the VAE reconstruction objective. After training, the network parameters are frozen. During validation, an independent clutter-only set is processed by the frozen model. Reconstruction deviation is evaluated as the detection statistic, and a decision threshold is determined by empirical quantile calibration for the prescribed false-alarm probability. During testing, each CUT is processed by the same frozen model, and its detection statistic is compared with the fixed threshold to obtain the final decision, without further training or threshold adjustment. This separation of background learning, threshold calibration, and target declaration enables unsupervised target detection in sea clutter with controllable false-alarm behavior.

\subsection{Complex-valued VAE}
VAEs are a class of deep generative models developed under the variational-inference framework \cite{kingma2013auto}. Let the observation and latent variables be denoted by \(\mathbf{x}\in\mathbb{R}^{D}\) and \(\mathbf{z}\in\mathbb{R}^{d}\), respectively. The joint distribution of the generative model is written as
\(p_{\theta}(\mathbf{x},\mathbf{z})=p_{\theta}(\mathbf{x}\mid\mathbf{z})p(\mathbf{z})\),
where the prior is typically specified as \(p(\mathbf{z})=\mathcal{N}(\mathbf{0},\mathbf{I})\).
Because the exact posterior \(p_{\theta}(\mathbf{z}\mid\mathbf{x})\) is generally intractable, an approximate posterior \(q_{\phi}(\mathbf{z}\mid\mathbf{x})\) is introduced. The parameters are then learned by maximizing the evidence lower bound (ELBO)
\begin{equation}\label{eq:elbo}
	\mathcal{L}(\theta,\phi;\mathbf{x})
	=\mathbb{E}_{q_{\phi}(\mathbf{z}\mid\mathbf{x})}\!\big[\log p_{\theta}(\mathbf{x}\mid\mathbf{z})\big]
	- D_{\mathrm{KL}}\!\big(q_{\phi}(\mathbf{z}\mid\mathbf{x})\,\|\,p(\mathbf{z})\big).
\end{equation}
A common choice is
\(q_{\phi}(\mathbf{z}\mid\mathbf{x})=\mathcal{N}\!\big(\boldsymbol{\mu}_{\phi},\operatorname{diag}(\boldsymbol{\sigma}^{2}_{\phi})\big)\),
for which differentiable sampling is enabled via the reparameterization trick,
\(\mathbf{z}=\boldsymbol{\mu}_{\phi}+\boldsymbol{\sigma}_{\phi}\odot\boldsymbol{\epsilon}\),
where \(\boldsymbol{\epsilon}\sim\mathcal{N}(\mathbf{0},\mathbf{I})\).

For radar baseband data, each slow-time sample contains in-phase and quadrature components. We use the notation \(\mathbf{x}\in\mathbb{C}^{L}\) for the received sequence and \(\mathbf{z}\in\mathbb{C}^{K}\) for the latent variable, while the implementation arranges each observation as paired I/Q channels. This notation provides a compact way to express the reconstruction likelihood and posterior model for the two-channel baseband sequence. Circularly symmetric complex Gaussian distributions \(\mathcal{CN}(\cdot)\) are then adopted:
\(p(\mathbf{z})=\mathcal{CN}(\mathbf{0},\mathbf{I})\),
\(p_{\theta}(\mathbf{x}\mid\mathbf{z})=\mathcal{CN}\!\big(f_{\theta}(\mathbf{z}),\sigma_x^{2}\mathbf{I}\big)\),
and
\(q_{\phi}(\mathbf{z}\mid\mathbf{x})=\mathcal{CN}\!\big(\boldsymbol{\mu}_{\phi},\operatorname{diag}(\boldsymbol{\sigma}^{2}_{\phi})\big)\),
where \(\boldsymbol{\mu}_{\phi}\in\mathbb{C}^{K}\) and \(\boldsymbol{\sigma}^{2}_{\phi}\in\mathbb{R}_{+}^{K}\) parameterize the variational posterior. The ELBO remains in the form of \eqref{eq:elbo}, while the KL divergence between complex Gaussian distributions yields
\begin{equation}\label{eq:ckl}
	D_{\mathrm{KL}}\!\big(q_{\phi}(\mathbf{z}\mid\mathbf{x})\,\|\,p(\mathbf{z})\big)
	=\sum_{k=1}^{K}\Big(
	-\log \sigma^{2}_{\phi,k}
	+\sigma^{2}_{\phi,k}
	+\big|\mu_{\phi,k}\big|^{2}
	-1\Big),
\end{equation}
in which \(|\cdot|\) denotes the complex modulus. Accordingly, the complex-valued reparameterization is expressed as
\(\mathbf{z}=\boldsymbol{\mu}_{\phi}+\boldsymbol{\sigma}_{\phi}\odot\boldsymbol{\epsilon}\),
with
\(\boldsymbol{\epsilon}=(\boldsymbol{\epsilon}_{r}+i\boldsymbol{\epsilon}_{i})/\sqrt{2}\),
and
\(\boldsymbol{\epsilon}_{r},\boldsymbol{\epsilon}_{i}\sim\mathcal{N}(\mathbf{0},\mathbf{I})\).
With this notation, the standard Gaussian reconstruction term can be replaced by a heavy-tailed observation likelihood while keeping the same VAE inference structure.

\subsection{Heavy-tailed Student-\(t\) reconstruction likelihood}
In clutter-only VAE training, a few spike-like sea-clutter samples can produce large reconstruction residuals and dominate a Gaussian reconstruction loss. This effect is undesirable because the learned background model should describe the prevailing clutter behavior rather than overfit occasional high-energy samples. To reduce the influence of such residuals, the decoder likelihood is modeled with a proper complex multivariate Student-\(t\) distribution \cite{johnson1995continuous}.

The probability density function is
\begin{equation}\label{eq:cts_pdf}
	p(\mathbf{x})
	= \frac{\Gamma(\nu+L)}{\Gamma(\nu)\,(\pi\nu)^{L}\det(\mathbf{\Sigma})}
	\left(1+\frac{1}{\nu}\delta(\mathbf{x})\right)^{-(\nu+L)},
\end{equation}
where \(\mathbf{x}\in\mathbb{C}^{L}\), \(\Gamma(\cdot)\) denotes the Gamma function, and
\(\delta(\mathbf{x})=(\mathbf{x}-\boldsymbol{\mu})^{H}\mathbf{\Sigma}^{-1}(\mathbf{x}-\boldsymbol{\mu})\).
The degrees of freedom \(\nu>0\) controls the tail thickness: smaller \(\nu\) gives heavier tails, whereas the Gaussian case is approached as \(\nu\to\infty\).

For the VAE decoder, \(\hat{\mathbf{x}}_{\theta}(\mathbf{z})\) is used as the location parameter, and \(\mathbf{\Sigma}_{x}\) describes residual dispersion. With the isotropic scale \(\mathbf{\Sigma}_{x}=\sigma_{x}^{2}\mathbf{I}\), the log-likelihood becomes
\begin{equation}\label{eq:cts_loglik}
	\begin{aligned}
		\ell_t(\mathbf{x},\mathbf{z})
		 & \triangleq \log p_{\theta}(\mathbf{x}\mid\mathbf{z}) \\
		 & = c -(\nu+L)\log\!\left(
		1+\frac{\big\|\mathbf{x}-\hat{\mathbf{x}}_{\theta}(\mathbf{z})\big\|_{2}^{2}}
		  {\nu\sigma_{x}^{2}}\right),
	\end{aligned}
\end{equation}
where
\(c=\log\Gamma(\nu+L)-\log\Gamma(\nu)-L\log(\pi\nu)-L\log\sigma_{x}^{2}\),
and \(\|\mathbf{u}\|_{2}^{2}\triangleq \mathbf{u}^{H}\mathbf{u}\).

Unlike a Gaussian loss, \eqref{eq:cts_loglik} grows slowly for large residuals. Let
\(\epsilon^{2}=\big\|\mathbf{x}-\hat{\mathbf{x}}_{\theta}(\mathbf{z})\big\|_{2}^{2}\),
the magnitude of the associated residual weight satisfies
\begin{equation}\label{eq:cts_weight}
	\left|
	\frac{\partial \ell_t(\mathbf{x},\mathbf{z})}{\partial \epsilon^{2}}
	\right|
	\propto \frac{\nu+L}{\nu\sigma_{x}^{2}+\epsilon^{2}},
\end{equation}
which shows that a sample with a larger residual receives a smaller effective weight in the reconstruction term.

The prior \(p(\mathbf{z})\) and variational posterior \(q_{\phi}(\mathbf{z}\mid\mathbf{x})\) remain circularly symmetric complex Gaussians. Replacing the Gaussian decoder likelihood by the Student-\(t\) likelihood gives
\begin{equation}\label{eq:elbo_ct}
	\begin{aligned}
		\mathcal{L}_{\mathrm{C},t}(\theta,\phi;\mathbf{x})
		 & =\mathbb{E}_{q_{\phi}(\mathbf{z}\mid\mathbf{x})}
		\!\big[\ell_t(\mathbf{x},\mathbf{z})\big]                                                   \\
		 & \quad- D_{\mathrm{KL}}\!\big(q_{\phi}(\mathbf{z}\mid\mathbf{x})\,\|\,p(\mathbf{z})\big).
	\end{aligned}
\end{equation}
Here, the first term is the heavy-tailed reconstruction criterion, and the second term is the latent regularization term.

\subsection{Time-domain amplitude reconstruction loss}

The Student-\(t\) likelihood reduces the influence of large complex residuals, but it does not separately control the magnitude sequence reconstructed over a dwell. For clutter-only training samples, preserving the observed slow-time magnitude fluctuations helps the decoder learn a background representation that is not overly smoothed. Therefore, an additional reconstruction term is imposed on the amplitude sequence.

Let the input be a slow-time complex vector \(\mathbf{x}\in\mathbb{C}^{L}\), let \(\mathbf{z}\sim q_{\phi}(\mathbf{z}\mid\mathbf{x})\), and let \(\hat{\mathbf{x}}=\hat{\mathbf{x}}_{\theta}(\mathbf{z})\in\mathbb{C}^{L}\) denote the decoder output. The amplitude sequences are defined as
\(\mathbf{a}\triangleq(|x_1|,\ldots,|x_L|)^{T}\) and
\(\hat{\mathbf{a}}\triangleq(|\hat{x}_1|,\ldots,|\hat{x}_L|)^{T}\),
where \(|\cdot|\) denotes the element-wise complex modulus. The amplitude reconstruction loss is
\begin{equation}\label{eq:amp_loss}
	L_{\mathrm{amp}}(\mathbf{x},\hat{\mathbf{x}})
	\triangleq \frac{1}{L}\big\||\mathbf{x}|-|\hat{\mathbf{x}}|\big\|_{2}^{2}
	=\frac{1}{L}\big\|\mathbf{a}-\hat{\mathbf{a}}\big\|_{2}^{2}.
\end{equation}
This term penalizes mismatch between the observed and reconstructed slow-time magnitude sequences. It complements the Student-\(t\) likelihood: the likelihood controls complex residuals with reduced sensitivity to large errors, whereas \(L_{\mathrm{amp}}\) directly constrains the magnitude behavior of clutter-only inputs.

\subsection{Training objective and detection statistic}
The training objective combines the Student-\(t\) ELBO with the amplitude reconstruction term:
\begin{equation}\label{eq:obj_def}
	\begin{aligned}
		\min_{\theta,\phi}\; J(\theta,\phi)
		 & =\mathbb{E}_{p_{\mathrm{data}}(\mathbf{x})}\!\Big[
			                                                -\mathcal{L}_{\mathrm{C},t}(\theta,\phi;\mathbf{x}) \\
			                                                &\qquad\qquad
			                                                +\lambda_{\mathrm{amp}}\,
			                                                \mathbb{E}_{q_{\phi}(\mathbf{z}\mid\mathbf{x})}
			                                                L_{\mathrm{amp}}\!\big(\mathbf{x},\hat{\mathbf{x}}_{\theta}(\mathbf{z})\big)
			                                                \Big],
	\end{aligned}
\end{equation}

Here, \(\lambda_{\mathrm{amp}}\ge 0\) controls the relative strength of the amplitude term. The objective therefore uses the Student-\(t\) term to model heavy-tailed complex residuals and the amplitude term to maintain slow-time magnitude consistency in the reconstructed clutter background.

After training, the network parameters are fixed and the reconstruction deviation is used for threshold-based detection. For an input \(\mathbf{x}\), let \(\hat{\mathbf{x}}\) be its reconstruction from the frozen model. The scalar detection statistic is defined as
\begin{equation}\label{eq:det_stat}
	T(\mathbf{x})
	=
	\frac{1}{L}\big\|\mathbf{x}-\hat{\mathbf{x}}\big\|_{2}^{2}
	+
	\lambda_{\mathrm{amp}}L_{\mathrm{amp}}(\mathbf{x},\hat{\mathbf{x}}).
\end{equation}
Given a prescribed false-alarm probability \(P_{fa}\), the threshold is calibrated from an independent clutter-only validation set:
\begin{equation}\label{eq:det_threshold}
	\eta
	=
	Q_{1-P_{fa}}\!\left(
	\left\{T(\mathbf{x}^{(0)}_n)\right\}_{n=1}^{N_{\mathrm{val}}}
	\right),
\end{equation}
where \(Q_{1-P_{fa}}(\cdot)\) denotes the empirical \((1-P_{fa})\)-quantile and \(\mathbf{x}^{(0)}_n\) is a validation sample under \(H_0\). The final decision is
\begin{equation}\label{eq:det_rule}
	T(\mathbf{x})
	\mathop{\gtrless}_{H_0}^{H_1}
	\eta .
\end{equation}

\subsection{Network implementation}

\begin{figure}[htbp]
	\centering
	\includegraphics[width=1\linewidth]{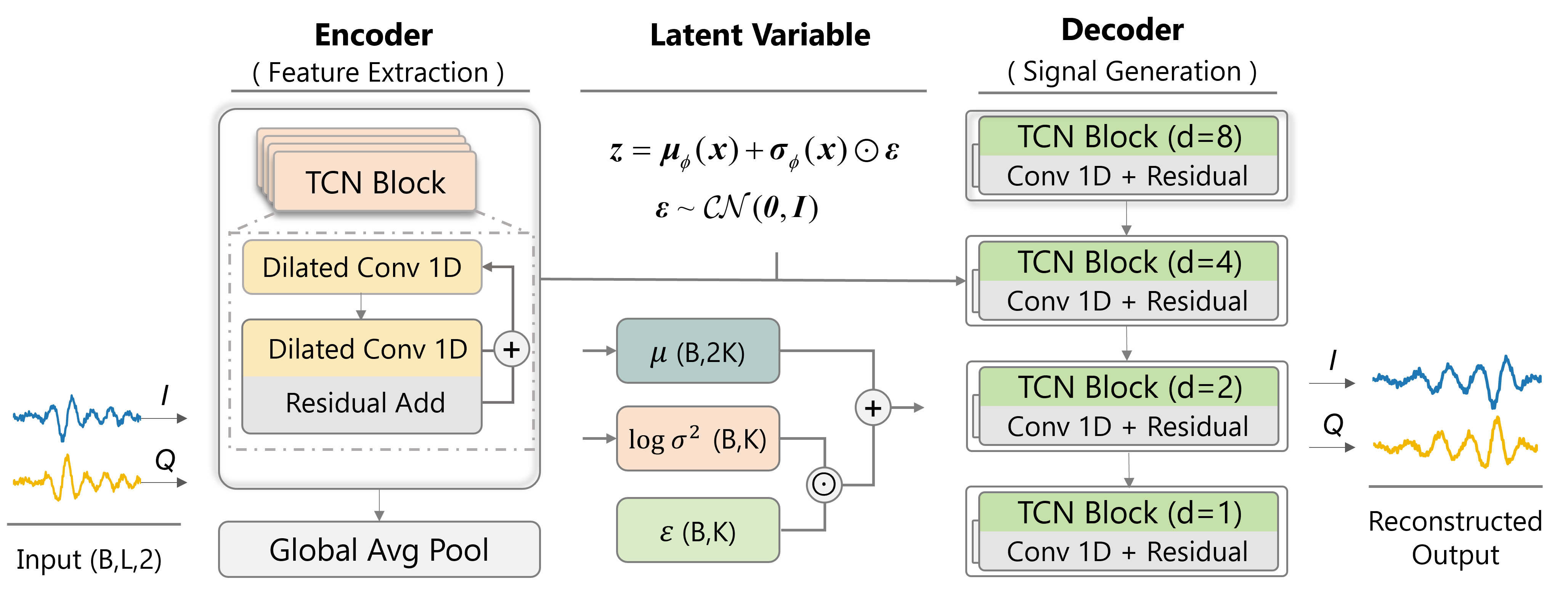}
	\caption{Architecture of the proposed complex VAE}
	\label{fig:vae_structure}
\end{figure}

The implemented VAE architecture is shown in Fig.~\ref{fig:vae_structure}. Each complex slow-time sequence \(\mathbf{x}\in\mathbb{C}^{L}\) is stored as its I/Q components, forming a two-channel tensor \(\mathbf{x}_{iq}\in\mathbb{R}^{L\times 2}\). The training set is arranged as \(\mathbf{X}\in\mathbb{R}^{N\times L\times 2}\), from which mini-batches \(\mathbf{X}_{b}\) are sampled. The network follows the standard VAE pipeline, consisting of an encoder, a reparameterization-based sampling module, and a decoder. Given \(\mathbf{x}_{iq}\), the encoder outputs a complex latent mean represented by separate real and imaginary heads, together with a positive real-valued scale vector. These parameters define \(q_{\phi}(\mathbf{z}\mid\mathbf{x})\), from which a latent vector \(\mathbf{z}\in\mathbb{C}^{K}\) is sampled and then mapped by the decoder to the I/Q reconstruction \(\hat{\mathbf{x}}_{iq}\).

Both encoder and decoder are implemented by stacking temporal convolutional (TCN) blocks, which model local dependencies along the slow-time dimension. During training, the decoder output is used in the Student-\(t\) reconstruction likelihood and in the amplitude reconstruction loss. Thus, the implementation combines I/Q reconstruction, heavy-tailed residual modeling, and slow-time magnitude regularization within the VAE framework.

\section{Experiments}

\subsection{Dataset description and preprocessing}
The evaluation uses an open set of measured sea-clutter records from \emph{Journal of Radars} \cite{Liu2019Sea-detecting}. We use the scanning-radar records acquired in Yantai, Shandong, China, by an X-band radar at sea state 3--4. Clutter samples for model training and validation are extracted from \path{20191012110735_10_scanning.mat}, while all test samples are taken from \path{20191012110757_19_scanning.mat}. The two files are kept separate throughout the experiments, so that threshold calibration and performance evaluation are not performed on the same recording.
\begin{figure}[htbp]
	\centering
	\begin{subfigure}[b]{0.7\linewidth}
		\centering
		\includegraphics[width=\linewidth]{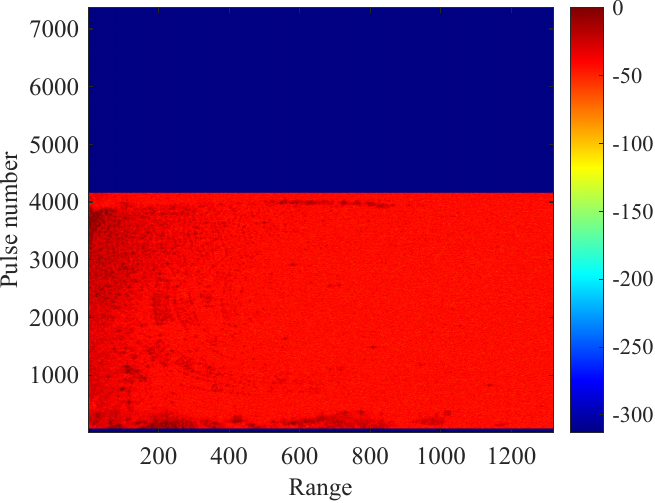}
		\caption{Time domain.}
		\label{subfig:fig_scan_time}
	\end{subfigure}
	\hfill
	\begin{subfigure}[b]{0.7\linewidth}
		\centering
		\includegraphics[width=\linewidth]{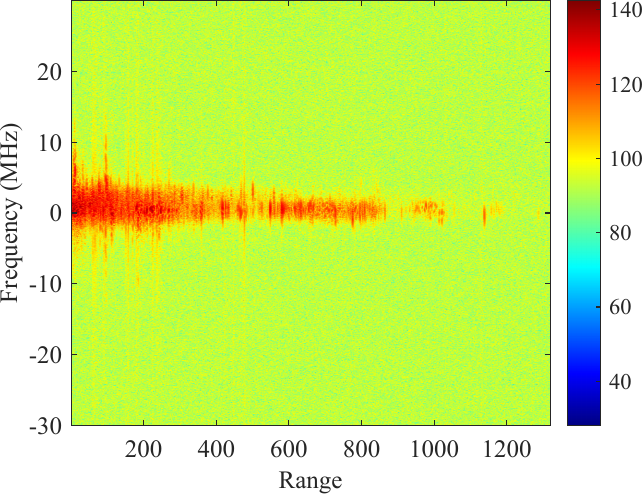}
		\caption{Doppler domain.}
		\label{subfig:fig_dataspt}
	\end{subfigure}
	\caption{Measured scanning sea-clutter records in the time and Doppler domains.}
	\label{fig:radar_scan}
\end{figure}

Each scan is stored as an \(M\times N\) complex data array, with \(M=7369\) pulses over one \(360^\circ\) scan and \(N=1320\) range cells. Since several azimuth sectors include strong land returns, as shown in Fig.~\ref{fig:radar_scan}, the experiments use the sea-dominated portion of each scan. Specifically, \(4000\) pulses are retained from each record and then segmented into slow-time windows for the proposed model and all comparison methods.

A fixed dwell length of \(L=16\) pulses is used for each slow-time input. For controlled SCNR evaluation on measured clutter, a synthesized point-target echo is added to the selected cell under test (CUT). The target amplitude is calibrated from the local background level: after excluding guard cells around the CUT, \(K_{\mathrm{ref}}=20\) neighboring reference cells are used to estimate the per-pulse clutter energy. For a prescribed \(\mathrm{SCNR}_{\mathrm{dB}}\), let \(\mathrm{SCNR}_{\mathrm{lin}}=10^{\mathrm{SCNR}_{\mathrm{dB}}/10}\), and set
\begin{equation}\label{eq:alpha_def}
	|\alpha|=\sqrt{\mathrm{SCNR}_{\mathrm{lin}}\,E_{b0}},
\end{equation}
where \(E_{b0}\) is the reference-cell estimate of the average clutter energy per pulse. The experiments sweep \(\mathrm{SCNR}\) from \(-5\) to \(10\,\mathrm{dB}\).

All network experiments were run in PyTorch with the same compact configuration: a four-layer 1D TCN encoder/decoder with \(32\) channels, latent size \(K=4\), Adam learning rate \(10^{-4}\), batch size \(32\), and \(50\) training epochs. At test time, Eq.~\eqref{eq:det_stat} provides one scalar score for each window. The threshold is fixed by the validation-set \(H_0\) scores in Eq.~\eqref{eq:det_threshold} before the independent test file is evaluated.

\subsection{Experimental results and analysis}
Detection performance was evaluated under a threshold fixed by validation data. For each \(\mathrm{SCNR}\) value, \(N_{\mathrm{MC}}=5000\) target-present trials were formed from the independent test recording by selecting an \(L\)-pulse slow-time window and inserting the synthesized target echo described above. The detection probability \(P_d\) and the realized false-alarm rate \(P_{fa}\) were then recorded under the fixed operating point. For each method, the threshold was determined before testing from clutter-only validation scores at the design false-alarm probability \(P_{fa}=0.01\). The compared methods included MF, AMF, \(\beta\)-VAE \cite{rouzoumka2025out}, and the proposed method.

\begin{figure}[htbp]
	\centering
	\includegraphics[width=0.7\linewidth]{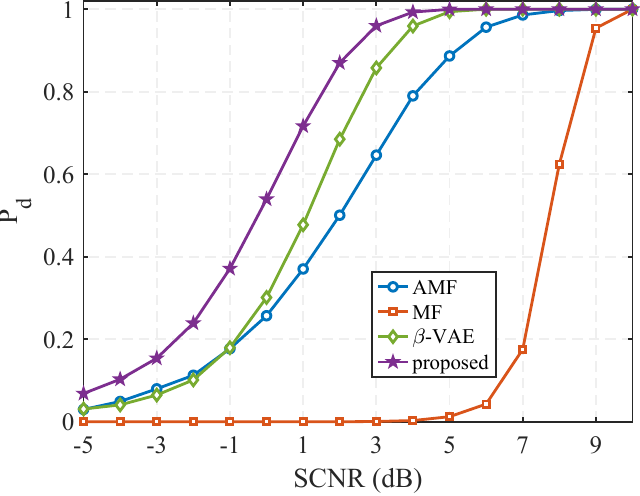}
	\caption{Detection probability versus SCNR for MF, AMF, \(\beta\)-VAE, and the proposed method under the design false-alarm probability \(P_{fa}=0.01\).}
	\label{fig:pd_snr}
\end{figure}

Figure~\ref{fig:pd_snr} presents the detection probability versus SCNR under the design false-alarm probability \(P_{fa}=0.01\). The proposed method gives a higher detection probability over the tested SCNR range and enters the rising part of the curve earlier than the baseline methods, indicating improved weak-target detectability under the same validation-based operating rule. Meanwhile, the realized \(P_{fa}\) of each method stays near \(0.01\), so the gain in \(P_d\) is not explained by a looser false-alarm operating point.

An ablation study was further conducted using the same data split, target-injection rule, and threshold-calibration procedure. The evaluated variants included a real-valued network baseline (Real net), an I/Q network baseline denoted as Complex net, and two extended variants that add the Student-\(t\) likelihood (\(+\,t\)-NLL) and the time-domain amplitude term (\(+\,\)time-amp).

\begin{figure}[htbp]
	\centering
	\includegraphics[width=0.7\linewidth]{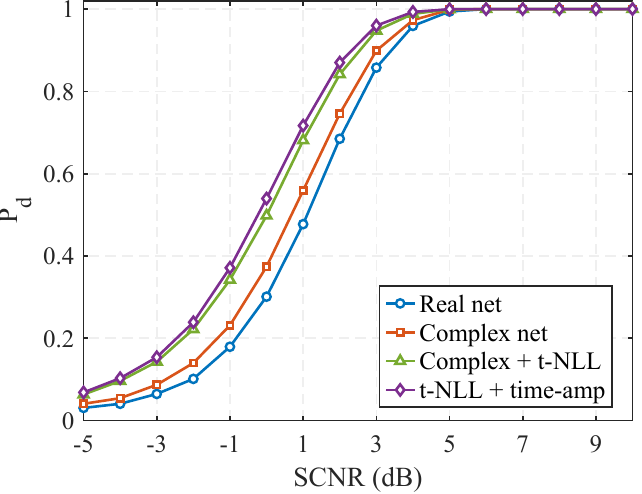}
	\caption{Detection probability versus SCNR for ablation variants under the design false-alarm probability \(P_{fa}=0.01\).}
	\label{fig:ablation}
\end{figure}

Figure~\ref{fig:ablation} shows the detection probability versus SCNR for these variants under the design false-alarm probability \(P_{fa}=0.01\). Adding the Student-\(t\) likelihood improves the operating curve relative to the I/Q baseline, and the amplitude term provides a further gain. This result indicates that the heavy-tailed residual model and the slow-time amplitude term both contribute to the final detector.

\begin{figure}[htbp]
	\centering
	\begin{subfigure}[b]{0.7\linewidth}
		\centering
		\includegraphics[width=\linewidth]{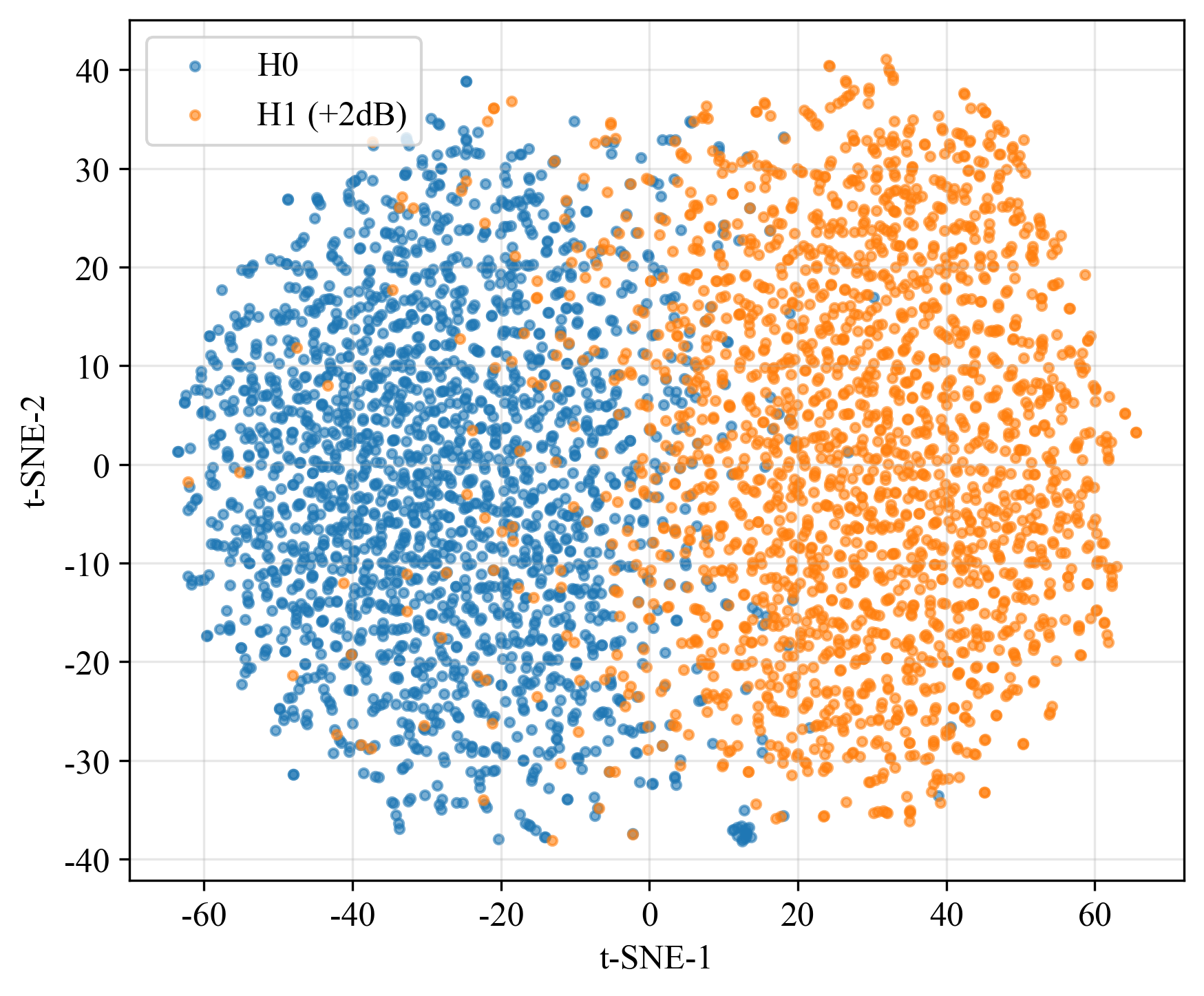}
		\caption{before}
	\end{subfigure}
	\hfill
	\begin{subfigure}[b]{0.7\linewidth}
		\centering
		\includegraphics[width=\linewidth]{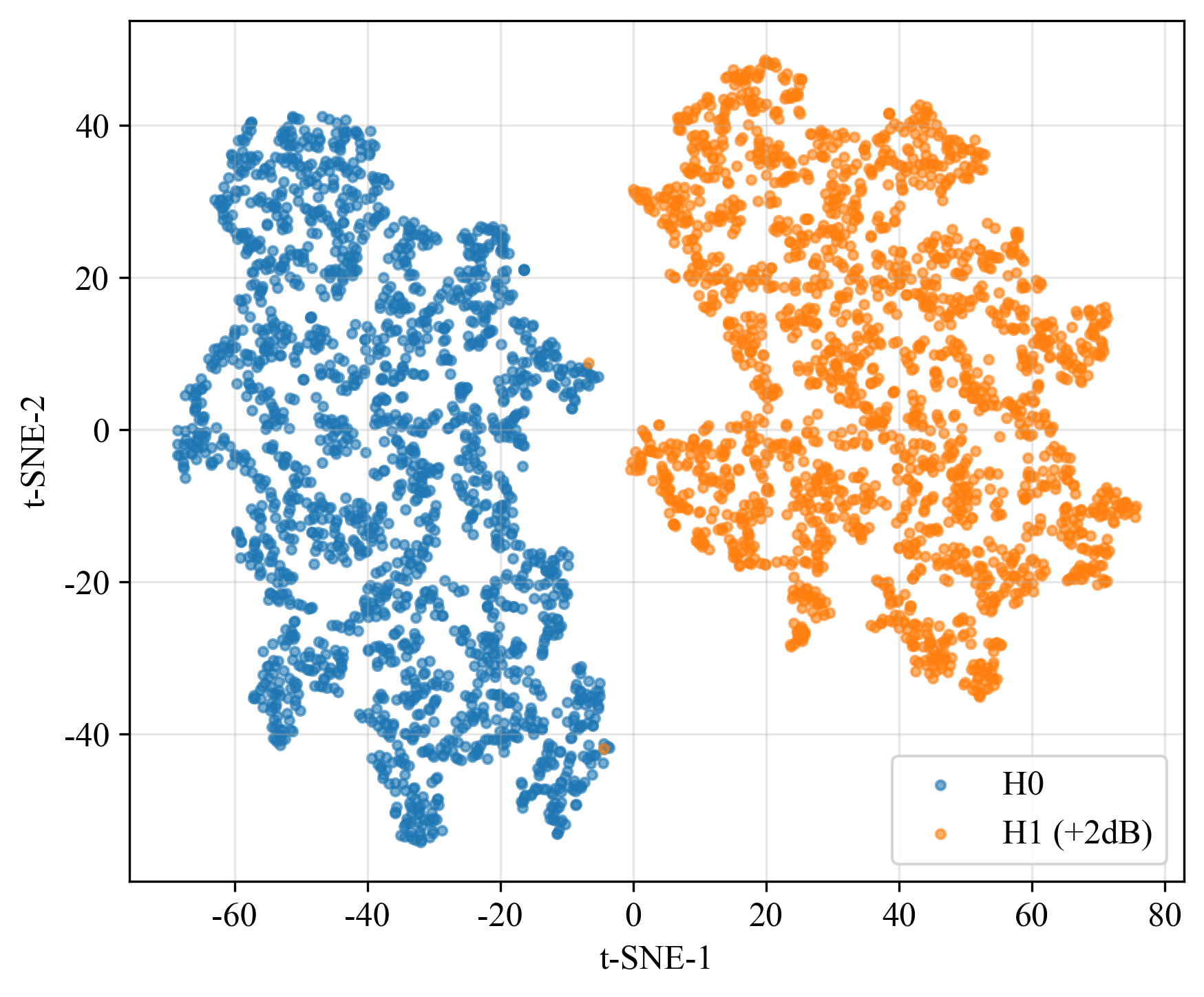}
		\caption{after}
	\end{subfigure}
	\caption{t-SNE maps of raw inputs and encoder-mapped features at \(\mathrm{SCNR}=2\,\mathrm{dB}\).}
	\label{fig:tsne}
\end{figure}

To provide a qualitative view of the learned feature space, Fig.~\ref{fig:tsne} reports t-SNE maps at \(\mathrm{SCNR}=2\,\mathrm{dB}\) before and after encoder mapping. Compared with the raw input view, the mapped features show a clearer distinction between clutter-only and target-containing samples. This visualization is used as supporting evidence for the detection curves rather than as an independent quantitative metric.

\section{Conclusion}
This paper presented an unsupervised VAE detector for weak maritime targets in measured sea clutter. Each baseband slow-time sequence is represented by paired I/Q channels, and the model combines a Student-\(t\) reconstruction likelihood with a slow-time amplitude reconstruction term. The resulting statistic is calibrated on clutter-only validation data and then applied to independent test recordings. Experiments show higher \(P_d\) than MF, AMF, and a real-valued \(\beta\)-VAE at the same design \(P_{fa}\), with the realized \(P_{fa}\) remaining close to the operating point. The ablation study indicates that the Student-\(t\) likelihood and the amplitude term both contribute to the final detector. Future work will examine threshold calibration and model robustness under more diverse sea states and measurement conditions.

\bibliographystyle{IEEEtran}
\bibliography{refs}

@article{2007An,
  title={An Adaptive Detection Algorithm},
  author={ Kelly, Edward J. },
  journal={IEEE Transactions on Aerospace \& Electronic Systems},
  volume={AES-22},
  number={2},
  pages={115-127},
  year={2007},
}

@article{fuhrmann1992cfar,
  title={A CFAR adaptive matched filter detector},
  author={Fuhrmann, Daniel R and Kelly, Edward J and Nitzberg, Ramon},
  journal={IEEE Trans. Aerosp. Electron. Syst},
  volume={28},
  number={1},
  pages={208--216},
  year={1992}
}

@article{jain2023radar,
  title={Radar clutter covariance estimation: A nonlinear spectral shrinkage approach},
  author={Jain, Shashwat and Krishnamurthy, Vikram and Rangaswamy, Muralidhar and Kang, Bosung and Gogineni, Sandeep},
  journal={IEEE Transactions on Aerospace and Electronic Systems},
  volume={59},
  number={6},
  pages={7640--7653},
  year={2023},
  publisher={IEEE}
}

@article{xue2021adaptive,
  title={Adaptive detection of radar targets in heavy-tailed sea clutter with lognormal texture},
  author={Xue, Jian and Liu, Jun and Xu, Shuwen and Pan, Meiyan},
  journal={IEEE Transactions on Geoscience and Remote Sensing},
  volume={60},
  pages={1--11},
  year={2021},
  publisher={IEEE}
}

@article{guo2024persymmetric,
  title={Persymmetric adaptive subspace detection in compound Gaussian sea clutter with generalized inverse Gaussian texture},
  author={Guo, Hongzhi and Wang, Zhihang and He, Zishu and Cheng, Ziyang},
  journal={Signal Processing},
  volume={216},
  pages={109300},
  year={2024},
  publisher={Elsevier}
}

@article{dong2012optimal,
  title={Optimal coherent radar detection in a K-distributed clutter environment},
  author={Dong, Y},
  journal={IET Radar, Sonar \& Navigation},
  volume={6},
  number={5},
  pages={283--292},
  year={2012},
  publisher={IET}
}

@article{wu2025adaptive,
  title={Adaptive radar target detection in nonzero-mean compound Gaussian sea clutter with random texture},
  author={Wu, Haoqi and Wang, Zhihang and Guo, Hongzhi and He, Zishu},
  journal={Signal Processing},
  volume={227},
  pages={109720},
  year={2025},
  publisher={Elsevier}
}

@article{jiang2023subspace,
  title={Subspace-based distributed target detection in compound-Gaussian clutter},
  author={Jiang, Qing and Wu, Yuntao and Liu, Weijian and Zheng, Daikun and Jian, Tao and Gong, Pengcheng},
  journal={Digital Signal Processing},
  volume={140},
  pages={104141},
  year={2023},
  publisher={Elsevier}
}

@article{qu2022false,
  title={A false alarm controllable detection method based on CNN for sea-surface small targets},
  author={Qu, Qizhe and Wang, Yong-Liang and Liu, Weijian and Li, Binbin},
  journal={IEEE Geoscience and Remote Sensing Letters},
  volume={19},
  pages={1--5},
  year={2022},
  publisher={IEEE}
}

@article{chen2021false,
  title={False-alarm-controllable radar detection for marine target based on multi features fusion via CNNs},
  author={Chen, Xiaolong and Su, Ningyuan and Huang, Yong and Guan, Jian},
  journal={IEEE Sensors Journal},
  volume={21},
  number={7},
  pages={9099--9111},
  year={2021},
  publisher={IEEE}
}

@article{wan2022sequence,
  title={Sequence-feature detection of small targets in sea clutter based on Bi-LSTM},
  author={Wan, Hao and Tian, Xiaoqing and Liang, Jing and Shen, Xiaofeng},
  journal={IEEE Transactions on Geoscience and Remote Sensing},
  volume={60},
  pages={1--11},
  year={2022},
  publisher={IEEE}
}

@article{wang2022maritime,
  title={Maritime radar target detection in sea clutter based on CNN with dual-perspective attention},
  author={Wang, Jingang and Li, Songbin},
  journal={IEEE Geoscience and Remote Sensing Letters},
  volume={20},
  pages={1--5},
  year={2022},
  publisher={IEEE}
}

@article{feintuch2023neural,
  title={Neural network-based multitarget detection within correlated heavy-tailed clutter},
  author={Feintuch, Stefan and Permuter, Haim H and Bilik, Igal and Tabrikian, Joseph},
  journal={IEEE Transactions on Aerospace and Electronic Systems},
  volume={59},
  number={5},
  pages={5684--5698},
  year={2023},
  publisher={IEEE}
}

@article{liang2022unsupervised,
  title={Unsupervised radar target detection under complex clutter background based on mixture variational autoencoder},
  author={Liang, Xueling and Chen, Bo and Chen, Wenchao and Wang, Penghui and Liu, Hongwei},
  journal={Remote Sensing},
  volume={14},
  number={18},
  pages={4449},
  year={2022},
  publisher={MDPI}
}

@article{ru2025optimizing,
  title={Optimizing latent space for effective radar target detection using variational auto-encoder},
  author={Ru, Hongtao and Xu, Shuwen and Zhang, Luxi and Shui, Penglang},
  journal={Signal Processing},
  pages={110235},
  year={2025},
  publisher={Elsevier}
}

@inproceedings{rouzoumka2025out,
  title={Out-of-Distribution Radar Detection in Compound Clutter and Thermal Noise through Variational Autoencoders},
  author={Rouzoumka, YA and Terreaux, E and Morisseau, C and Ovarlez, J-P and Ren, C},
  booktitle={ICASSP 2025-2025 IEEE International Conference on Acoustics, Speech and Signal Processing (ICASSP)},
  pages={1--5},
  year={2025},
  organization={IEEE}
}

@article{akrami2019robust,
  title={Robust variational autoencoder},
  author={Akrami, Haleh and Joshi, Anand A and Li, Jian and Aydore, Sergul and Leahy, Richard M},
  journal={arXiv preprint arXiv:1905.09961},
  year={2019}
}

@article{Liu2019Sea-detecting,
  title={Sea-detecting X-band radar and data acquisition program},
  author={Liu, Ningbo and Dong, Yunlong and Wang, Guoqing and others},
  journal={Journal of Radars},
  volume={8},
  number={5},
  pages={656--667},
  year={2019},
  doi={10.12000/JR19089}
}

@article{kingma2013auto,
  title={Auto-encoding variational bayes},
  author={Kingma, Diederik P and Welling, Max},
  journal={arXiv preprint arXiv:1312.6114},
  year={2013}
}

@book{johnson1995continuous,
  title={Continuous univariate distributions, volume 2},
  author={Johnson, Norman L and Kotz, Samuel and Balakrishnan, Narayanaswamy},
  volume={2},
  year={1995},
  publisher={John wiley \& sons}
}

\end{document}